
\raggedright
\tolerance=100000
\hyphenpenalty=1000
\raggedbottom
\font\twelvebf=cmbx12

\def\ninit{\hoffset=.0 truecm
            \hsize=16. truecm
            \vsize=22.1 truecm
            \baselineskip=19.pt
            \lineskip=0pt
            \lineskiplimit=0pt}

\def\pag{\pageno=2\footline={\hss\tenrm\folio\hss}}
\ninit
%
%

\def\mic{\,\mu{\rm m}}

\def\Msol{\,{\rm m_\odot}}
\def\Zsol{\,{\rm Z_\odot}}
\def\Lsol{\,{\rm L_\odot}}
\def\Msyr{\,{{\rm m_\odot}\,{\rm yr}^{-1}}}
%
%

\def\lsim{\,\lower2truept\hbox{${< \atop\hbox{\raise4truept\hbox{$\sim$}}}$}\,}
\def\gsim{\,\lower2truept\hbox{${> \atop\hbox{\raise4truept\hbox{$\sim$}}}$}\,}

%
%
\def\oneskip{\vskip\baselineskip}
\centerline{\null}
\nopagenumbers
\oneskip
\noindent
\centerline{\twelvebf Far-IR properties of early type galaxies}

\oneskip
\oneskip
\parindent=1truecm
\parskip 0pt

\centerline{P. Mazzei${}^1$, and G. De Zotti${}^1$}

\oneskip
\noindent
\llap{${}^1$ }{\it Osservatorio Astronomico, Vicolo dell'Osservatorio 5,
I--35122 Padova, Italy}

\oneskip
\oneskip
\oneskip
\noindent
\settabs 2 \columns
\+ Address for correspondence: &Dr. Gianfranco De Zotti\cr
\+&Osservatorio Astronomico\cr
\+&Vicolo dell'Osservatorio, 5\cr
\+&I--35122 Padova (Italy)\cr

\oneskip
\noindent

\vfill\eject

\pag
\centerline{ABSTRACT}

\smallskip\noindent
We have investigated the IRAS colours and the far-IR to optical luminosity
ratios of a complete sample of elliptical's and S0's brighter than
$B_T = 12$. On the average, elliptical galaxies
emit in the far-IR less than 1\% of their bolometric luminosity, while
S0's are about a factor of three brighter in the far-IR. There is a
considerable spread in the far-IR properties of individual galaxies.
On the average, the photospheric
emission of red giant stars can account for 50--60\% of the
$12\mic$ flux from early type galaxies; the contribution from diffuse
dust at this wavelength is $< 10\%$ in the case of ellipticals, and
may amount to 20--40\% for S0's. An additional, $\sim 30--40\%$, contribution
from circumstellar emission from evolved giants with mass loss (particularly
OH/IR stars) seems to be required in the case of ellipticals.
This suggests a small but significant star formation activity in
these galaxies at a look-back
time of 1--$2\,$Gyr, corresponding to about 10\% of that typical of
a disk galaxy having the same V-band luminosity. As for S0's, the
larger diffuse dust emission may swamp to some extent that of circumstellar
dust, which is indicated to be, on the average, $\lsim 20\%$.
The weak emission from diffuse
interstellar dust, detected mostly at $60\mic$ and $100\mic$, has color
temperatures similar to those of disk galaxies; as in that case of the latter,
a warm dust component is suggested, associated to star-forming regions.
The implied star formation would be a few percent of that of disk galaxies
of similar V-band luminosity and
could account for a fraction of the observed
UV branch of early type galaxies.

\medskip\noindent
{\it Subject headings:\/} galaxies: elliptical and lenticular -- infrared:
galaxies -- galaxies: interstellar medium

\vfill\eject

\centerline{1.\  INTRODUCTION}

\medskip
Far-IR properties of early-type systems are still poorly known because they
were only weakly detected by IRAS. On the other hand, far-IR data provide
important information on the interstellar medium as well as on star formation
activity (Thronson \& Bally 1987), and on mass loss from stars (Soifer et al.
1986; Jura, Kim \& Guhathakurta 1987; Knapp, Gunn, \& Wynn-Williams 1992).

Co-added IRAS data on a large sample of early type galaxies ($\sim 1150$)
have been presented by Knapp, Gunn, \& Wynn-Williams (1989).
As stressed by the authors, however, the
sample is not complete in any sense.
An even more delicate problem, also discussed by Knapp et al. (1989),
is the morphological classification, which is often very
uncertain; in fact spirals with faint disks may be misclassified as
S0's or even ellipticals, while elliptical with dust patches may be
classified as I0, S0 or even spiral galaxies. In order to have a
morphological classification as uniform and reliable as possible we confined
ourselves to galaxies with $B_T \leq 12$, classified E and S0
in the second edition of the Revised Shapley-Ames catalog of Bright
Galaxies (RSA; Sandage \& Tammann 1987). The chosen limiting magnitude is
that where incompleteness sets in, according to Sandage \& Tammann (1987).
IRAS fluxes or upper limits for almost all galaxies in the sample,
namely for 47 E and 60 S0 galaxies, are given by Knapp et al. (1989).
For ``large'' galaxies we have used the IRAS fluxes given in Table 7
of Knapp et al. (1989).


The IRAS detection rate for galaxies in the sample is anyway low. Therefore
to obtain meaningful results, upper limits must be taken into
account. To this end, we have exploited the survival analysis techniques
(Feigelson \& Nelson 1985; Schmitt 1985).

The plan of the paper is the following. In Sect. 2 we estimate the
far-IR to optical luminosity ratios and the far-IR colours for elliptical and
S0 galaxies. In Sect. 3 we summarize our model and discuss the far--IR colours.
The main results are presented in Sect. 4.

\bigskip
\centerline{2.\  FAR-IR TO OPTICAL LUMINOSITY RATIOS AND FAR-IR COLORS}

\medskip
Figures 1 and 2 show the distributions of logarithms of
far-IR to $B_T(0)$ flux density ratios
of E and S0 galaxies, respectively, for the four IRAS bands,
reconstructed exploiting the Kaplan-Meyer estimator, taking into
account upper limits. IRAS fluxes at 25, 60, and $100\mic$
have been taken from Knapp et al. (1989).

On the other hand,
Knapp et al. (1992) found that the $12\mic$ emission
is extended on the scale of the galaxy and pointed out that the IRAS
equivalent point source flux at this wavelength underestimates the
total flux. Corrected $12\mic$ fluxes or upper limits are given by these
authors for 30 elliptical galaxies in our sample. The mean
fraction of the total flux registered by the IRAS point-source fitting
procedure for these galaxies has been used to correct the fluxes of
the remaining 17 ellipticals and of S0's.
As for the other IRAS bands, we assume
that no correction of this kind is required since the emission is mostly
due to dust (see below) which is likely to be much more centrally
concentrated. Should a significant amount of dust be present in the
outer parts of early type galaxies, its temperature would be anyway
too low (due to the low radiation field intensity) to yield an
important contribution in the IRAS bands (see below).

B-band
fluxes are referred to the effective wavelength $\lambda = 0.44\,\mu$m;
the calibration given by Johnson (1966) was adopted.

The use of logarithmic values minimizes the effect on the estimated means
and variances of anomalous values, that may correspond to
misclassified objects. Using the ratios directly leads
to estimated mean values which are sistematically higher by a factor 2--3.
Nevertheless, particularly when the fraction of detected objects is low,
the reconstructed distribution has a large peak at the lowest observed ratio,
i.e. the low side of the distribution is strongly curtailed, again with the
consequence of an overestimation of the mean. A safer estimator of the
true average value is, in this case, the median, and we have adopted it;
the error has been estimated from the width of the distribution above
the median value. The results are listed in Table 1.

How crucial
a reliable morphological classification is in the present
context, is illustrated by Figure 3 which shows the distributions of
far-IR to $B_T$ flux ratios for galaxies classified as ellipticals
in Table 2 of Knapp et al. (1989). The tails extending to high values
of flux ratios are most likely due to misclassified S0 or spiral galaxies,
as indicated by the fact that they largely disappear if we confine ourselves
to the brightest galaxies and we adopt the morphological classification
given in the Third Reference
Catalogue of Bright Galaxies (de Vaucouleurs et al. 1991).

The far-IR emission is obviously very small in comparison with the optical.
The total flux between $42.5\,\mu$m
and $125.5\,\mu$m, $F_{\rm FIR}$, can be estimated
using the formula $\log (F_{\rm FIR})
= \log [1.26\times(F_{60} + F_{100})]$ given in {\it Cataloged Galaxies
and Quasars Observed in the IRAS Survey} (1989), where fluxes are in
$\hbox{W}\,\hbox{m}^{-2}$, $F_{60} = 2.58\, 10^{-14} f_{60}$,
$F_{100} = 1.00\, 10^{-14} f_{100}$, with $f_{60}$ and $f_{100}$ measured
in Jy. Setting $F_B = (\lambda f_\lambda)_B$, the ratio
$F_{\rm FIR}/F_B$ can be derived either from the ratios $f_{60}/f_{100}$
and $f_{100}/f_{B}$ or from $f_{100}/f_{60}$ and $f_{60}/f_{B}$.
The results are similar in both cases. We find $F_{\rm FIR}/F_B
\simeq 0.007$ for ellipticals  and $F_{\rm FIR}/F_B \simeq 0.02$ for S0s, i.e.
the far-IR emission of S0s turns out to be significantly
higher than that of ellipticals with the same $B_T$.

Note that if the dust distribution follows that of stars, we may expect
a large dust mass in the outer regions of ellipticals. Such dust would
feel a low intensity radiation field and would thus be very cold. Its
contribution to flux in the IRAS bands could be small even if its
global emission isn't. It may then be possible that the global
dust emission is significantly or even much larger than $F_{\rm FIR}$
(see \S 3.2).

The distributions of far-IR colors of E and S0 galaxies
are shown in Figs. 4 and 5, respectively; the median values are given in
Table 2. The most conspicuous difference
with the far-IR spectrum of nearby disk galaxies (Hawarden et al. 1986;
Puxley, Hawarden, \& Mountain 1988; Xu \& De Zotti 1989) are
the substantially higher
ratios of $12\mic$ and $25\mic$ to $60\mic$ or $100\mic$ fluxes, particularly
in the case of ellipticals (see Fig. 6). The $60\mic/100\mic$ ratios of
both E and S0 galaxies are very close to those found for nearby disk galaxies.

\bigskip\bigskip
\centerline{3.\  DISCUSSION}

\medskip
\centerline{\it 3.1. $12\mic$ emission}

\medskip\noindent
The most direct indicator of possible contributions at $12\mic$ in addition
to the photosheric emission of cool stars, which dominate at $2.2\mic$,
is obviously the ratio of $12\mic$ to $2.2\mic$ flux densities.

We have adopted the total fluxes at $2.2\mic$, referred to the effective
aperture of the IRAS $12\mic$ beam, given by Knapp et al. (1992)
for the 30 E galaxies in our sample for which these data are available.
For 28 out of the 60 S0's, we have used the
K magnitudes corrected to an aperture $\log A/D_0 = 0$ and for
extinction within our own Galaxy listed by Marsiaj (1992).
For the remaining galaxies, the K magnitude
has been derived from the B magnitude assuming $B-K = 4.1$ (Aaronson
1978; de Vaucouleurs \& de Vaucouleurs 1972). The $2.2\mic$ flux was
computed using Johnson's (1966) calibration.

Our analysis yields a median value $f_{12}/f_{2.2}\simeq 0.145 \pm 0.02$
for ellipticals and $0.14 \pm 0.02$ for S0s, close to the mean value
$\langle f_{12}/f_{2.2}\rangle = 0.138 \pm 0.014$, found by Knapp
et al. (1992), based on a slightly more heterogeneous sample. There is
however a considerable dispersion, as shown by Figure~5. For
comparison, for the bulge of M31 such ratio is $f_{12}/f_{2.2} \simeq 0.10$
within an aperture of diameter $4'$ centered on the nucleus (Soifer et al.
1986; Knapp et al. 1992).

An important contribution to the $12\mic$ luminosity
comes from direct photospheric emission of cool stars, mostly red
giants, which also account for the bulk of the $2.2\mic$ luminosity.
The $12\mic/2.2\mic$ luminosity ratio of these stars increases substantially
with metallicity, as expected since the giant branch is cooler for
metal rich objects. For solar metallicity,
from the $V- [12\mic]$ colors reported by Waters, Cot\'e, \& Aumann (1987),
coupled with the $V-K$ colours given by Johnson (1966), we find
$12\mic/2.2\mic$ flux density ratios, $f_{12}/f_{2.2}$, ranging from
$\simeq 0.045$ for K-giants to 0.06 for M-giants.
Using the $V-K$ colors
given by Frogel \& Whitford (1987) and the $V- [12\mic]$ colors
by Waters et al. (1987) we have also estimated
the mean $12/2.2\mic$ flux density ratio for the super metal rich galactic
bulge M giants (spectral types M1--M5); we find
$\langle f_{12}/f_{2.2}\rangle \simeq 0.135$.
Knapp et al. (1992) have analyzed two samples of evolved stars to find
that most of them have $\langle f_{12}/f_{2.2}\rangle \simeq 0.08$.

On account of the fact that metallicities of ellipticals appear to decrease
from 2--$3\Zsol$ at the center to $\Zsol$ at one half-light radius (Munn 1992),
we guess that the effective metallicity is unlikely to be much different
from $\Zsol$ and, following Knapp et al. (1992), we adopt 0.08 as the
reference value for the mean $f_{12}/f_{2.2}$ ratio for stellar photospheres.

Figure~5 shows that most early type galaxies
have a substantial $12\mic$ excess over photospheric emission. If
$\langle f_{12}/f_{2.2}\rangle_{\rm photosph} = 0.08$, such
excess amounts, on the average, to $\sim 40\hbox{--}50\%$ (however with
a large uncertainty).

The contribution of diffuse dust can be estimated from the $100\mic$ flux,
assuming a $f_{12}/f_{100}$ ratio similar to that observed for
cold dust in disk galaxies (the ``warm'' dust component, associated to
regions of higher radiation field intensity such as HII regions,
has substantially lower  $f_{12}/f_{100}$ ratios, due to the destruction
of PAH molecules, held responsible for the bulk of the $12\mic$ flux
in disk galaxies, by UV photons and by interactions with ionized gas;
cf. Xu \& De Zotti 1989).
The median value of this ratio observed for the sample of
$100\mic$ cirrus studied by Paley et al. (1991) is $\simeq 0.035$;
for the sample of nearby disk galaxies defined by Hawarden et al. (1986)
we get $\langle f_{12}/f_{100}\rangle \simeq 0.024$; for spirals
in the Virgo cluster, surveyed by Helou et al. (1988), Knapp et al. (1992)
find $\langle f_{12}/f_{100}\rangle \simeq 0.042$. Taking into account
that (see Table 2) for ellipticals the median $f_{12}/f_{100}\simeq
0.5$ (taking into account upper limits),
we infer that the contribution of diffuse dust to the $12\mic$ flux of
these galaxies is less than $10\%$.

In the case of S0's the median $f_{12}/f_{100}$ ratio is
$\simeq 0.1$ (Table 2), so that diffuse dust
can contribute $\sim 20$--40\% of the mean $12\mic$ flux.

A detailed analysis of IRAS
observations of the nuclear bulge of M31, the closest example of an
old stellar population similar to that of ellipticals, indicates
(Soifer et al. 1986) that one further important contributor
at $12\mic$ is hot circumstellar dust
(see also Impey, Wynn-Williams \& Becklin 1986), primarily associated
to OH/IR stars (Soifer et al. 1986; Cox, Kr\"ugel \&
Mezger 1986). Knapp et al. (1992) demonstrated that
the $12\mic$ emission is distributed like the starlight and is
therefore likely due to photospheric and circumstellar emission from
red giant stars. The present results suggest that circumstellar dust
may provide, on the average, as much as
$\simeq 30$--40\% of the $12\,\mic$ luminosity of ellipticals
and $\lsim 20$\% of that of S0's.

OH/IR stars are believed to be in the final stage of evolution along the
asymptotic giant branch (AGB). Their pulsation and kinematic properties are
consistent
with initial masses in the range $m_{HeF} \leq m \leq m_{up}$
($m_{HeF}$ being the maximum mass for
violent ignition of helium burning and $m_{up}$ the  maximum
mass of stars that develop a highly degenerate CO core), i.e. in the range
where stars experience a prolonged, thermally pulsating AGB phase (Engels
et al. 1983; Baud et al. 1981; Feast 1963).

Following Mazzei, Xu \& De Zotti (1992),
we assume the contribution of OH/IR stars to the emission of a galaxy
at each galactic age $T$ to be a fixed fraction $F$ of the global
bolometric luminosity of AGB stars with initial masses in the quoted range,
and we adopt the spectrum of OH~27.2+0.2 (Baud et al. 1985)
as representative for stars of this class (see also Cox \&
Mezger 1989). Then, the total luminosity of OH/IR stars in the passband
$\Delta\lambda$ is given by:
$$ L_{OH,\Delta\lambda} (T) = F \int _{T_{AGB}(m_{up})}^T d\tau\,\psi(T-\tau)
\int _{m_{l,OH}(\tau)}^{m_{u,OH}(\tau)} \phi (m)
10^{-0.4 (M_{\Delta \lambda} (m, \tau) - M_\odot)} dm\ \ \Lsol , \eqno(1)$$
where $\psi(t)$ is the star formation rate (SFR), $\phi(m)$ is the
initial mass function (IMF),
${T_{AGB}(m_{up})}$ is the time when the first OH/IR
stars appear, $m_{l,OH}(\tau)\ (\geq m_{HeF})$ and
$m_{u,OH}(\tau)\  (\leq m_{up})$ are the minimum and
the maximum mass of OH/IR stars of age $\tau$.
The coefficient $F$ was determined by Mazzei et al. (1992)
from the condition that
OH/IR stars account for 10\% of the observed $12\mic$ luminosity of our
Galaxy (Ghosh, Drapatz \& Peppel 1986;
Boulanger \& P\'erault 1988). They find
$F=0.05$, in good agreement with Herman \& Habing's (1985) estimate
based on a different approach.

Due to the steep increase of the $M/L$ ratio with mass, most of the
contribution comes from stars at the upper mass limit, whose lifetime
is $\sim 1\hbox{--}2\,$Gyr. Since, moreover, the SFR
is believed to be a rapidly (exponentially)
decreasing function of galactic age (for early type galaxies), the
contributon of OH/IR stars to the
present $12\mic$ luminosity should essentially reflect the SFR 1--$2\,$Gyr
ago.

If OH/IR stars are to account for, say, 30\% of the $12\mic$ luminosity of
early type galaxies,
the SFR at that time should be about 10\% of that of
a disk of equal mass, age and IMF, i.e. about $0.3\Msyr$ for
a total mass of $10^{11}\Msol$. From
the contribution of circumstellar dust to the $12\mic$ luminosity we may
derive, following Jura et al. (1987), an estimate of the total mass loss
from evolved stars, which is roughly given by $\dot M (\Msyr) = 0.045
(L_{12,{\rm circumstellar}}/10^8\Lsol)$, corresponding to typical
values $\sim 0.2\Msyr$.

\medskip
\centerline{\it 3.2. Diffuse dust emission}

\medskip\noindent
The diffuse dust emission spectrum is modelled
following Xu \& De~Zotti (1989), i.e. taking into account
the contributions of two components: warm dust, located in regions
of high radiation intensity (e.g., in the neighborhood of OB clusters)
and cold dust, heated by the general interstellar radiation field.
The model allows for a realistic grain--size distribution and includes
PAH molecules. The emission spectrum of PAH's is computed following
Puget, L\'eger, \& Boulanger (1985) and using the absorption cross sections
and the bandwidths at 3.3, 6.2, 7.7, 8.6 and $11.3\mic$ given
in Table 2 of Puget \& L\'eger (1989).
We have reckoned the  contributions of PAH in the IRAS
$12\mic$ band which encompasses the emission features at $7.7\mic$,
$8.6\mic$, and $11.3\mic$, and in the ISOCAM filters LW2 and SW1,
covering the wavelength ranges 5--$8.5\mic$ and 3.05--$4.1\mic$,
respectively.

The temperature distribution of cold dust has been computed exploiting
Jura's (1982) model, which assumes a
simple, spherically symmetric, stellar density distribution,
matching reasonably well King's (1966) model.
The adopted central radiation field intensity,
$16\,\hbox{V~mag}\,\hbox{arcsec}^{-2}$ corresponds to $I_0=46 I_
{\rm local}$ (see Xu \& De Zotti 1989); the core radius $r_0$ is taken to
be 100 pc.

The dust is assumed to be uniformly distributed up to a galactocentric radius
$r_d$. The dust
temperature drops quickly with increasing radius following
the decrease in the radiation intensity. Thus the dust temperature distribution
peaks at lower and lower temperatures as $r_d$ increases.
For example, if $r_d=30r_0$ the cold dust emission
peaks around $100\mic$ while for $r_d=100 r_0$ it peaks around  $150\mic$.

A fit of the average ratios $f_{25\mic}/f_{60\mic}$ and
$f_{60\mic}/f_{100\mic}$ listed in Table 2
implies, both for ellipticals and S0s, $I_w=110\,I_{\rm local}$ if
$r_d=30r_0$ or
$I_w=90\,I_{\rm local}$ if $r_d=100r_0$, and
a warm to cold luminosity ratio, $R_{w/c}$, of $0.53$ or $0.27$
respectively.

As already mentioned, the warm dust component is most easily understood as
associated to star formation regions, so that the ratio $R_{w/c}$ is
in some sense a measure of the star formation rate. The observed far-IR
spectrum of E and S0 galaxies might then provide some indication of
a significant residual star formation. Pursuing the
argument a little further, it is possible to quantify it, albeit crudely.

The estimated warm dust emission is
in the range 0.1--0.3\% of the bolometric luminosity for ellipticals,
and $\simeq 0.4\hbox{--}0.5\%$ for S0's. According to Leisawitz
\& Hauser (1988)  roughly 37\% of the total radiated energy from an
OB cluster is absorbed by nearby (and hence warm) dust grains, suggesting
that OB stars account for $\sim 0.3$--1\% of the bolometric luminosity of
early type galaxies. For comparison, the observed UV (0.12--$0.36\mic$)
luminosity of early type galaxies, ranges from 1.1 to 1.7\% of
the bolometric luminosity (Burstein et al. 1988). The above argument may
suggest that a significant fraction of the
UV flux could be accounted for by recent star formation; the required
SFR would be a few percent of that of a disk of equal
bolometric luminosity
and IMF.

\bigskip\bigskip
\centerline{4.\  CONCLUSIONS}

\medskip\noindent
Only a small fraction of the bolometric emission of early type galaxies comes
out in the far-IR. The blue to far-IR luminosity ratio,
$(\lambda L_\lambda)_B/L_{\rm FIR}$ ($L_{\rm FIR}$ being the luminosity
in the 42.5--$125.5\mic$ range) is only
$\simeq 0.7\%$ for ellipticals and $\simeq 2\%$ for S0's. Integrating over
a typical spectral energy distribution of early type galaxies, we find
that the bolometric luminosity is about $4.2 (\lambda L_\lambda)_B$. The
global far-IR luminosity is made uncertain by the lack of sub-mm data;
correspondingly the long wavelength emission from very cold dust
is essentially unconstrained. Models discussed here imply $L_{\rm dust, tot}
\simeq 2.6 \hbox{--} 4.8 L_{\rm FIR}$, depending on the extent of the
dust distribution. Thus, it is likely that dust absorbs and reprocesses
only less than 1\% of starlight of ellipticals and at most a few
percent of that of S0's, to be compared
with a typical 30\% in the case of normal disk galaxies.

Of particular interest are data on the $12\mic$ emission. As extensively
discussed by several authors (Impey et al. 1986; Knapp et al. 1992), the
observed $12\mic$ flux from elliptical galaxies is in excess of that expected
from direct photospheric emission of red giants and from diffuse dust
(or, rather, PAH molecules). A significant contribution from hot
circumstellar dust, surrounding evolved, mass losing, red giants
(particularly OH/IR stars) is indicated. Its amplitude, however, is difficult
to ascertain; the present analysis suggests
$\sim 30\%$. We have argued
that this mostly comes from stars with mass close to $m_{up}$,
the maximum mass of stars that develop a highly degenerate CO core,
whose lifetimes are $\sim 1\hbox{--}2\,$Gyr. This implies a small,
but significant, star formation activity in ellipticals during the last
few Gyrs: assuming an exponentially decreasing SFR, with a timescale
$\simeq 1\,$Gyr, we find that
if OH/IR stars are to account for $\simeq 30\%$ of the
$12\mic$ luminosity, the SFR at a look-back time of $2\,$Gyr had to be
$\simeq 10\%$ of that of a disk of equal V-band luminosity. This implies that
the photometric evolution of early type galaxies was not purely passive.

An additional, albeit weak, evidence that some star formation may be still
taking place in present day early type galaxies comes from the fact that the
mean $60/100\mic$ color temperature is very close to that of disk galaxies,
suggesting, as in the case of the latter galaxies, a significant warm
dust component associated to star-forming regions. A star formation
of a few percent of that appropriate to a disk galaxy having the same
V-band luminosity is indicated (see also Thronson \& Bally 1987); such star
formation could account for a fraction of the observed
UV branch of early type galaxies.

\bigskip\noindent
{\it Acknowledgements.} We are indebted to C. Xu for having worked out
the calculations of the diffuse dust emission spectrum,
and to L. Danese and A. Franceschini for valuable discussions. Work supported
in part by ASI and by the EEC programme ``Human Capital and Mobility''.

\vfill\eject

%
%
%

\def\aa #1 #2{{A\&A,}~{#1}, {#2}}
\def\aas #1 #2{{A\&AS,}~{#1}, {#2}}
\def\aar #1 #2{{A\&A Rev,}~{#1}, {#2}}
\def\advaa #1 #2{{Adv. Astron. Astrophys.,}~{#1}, {#2}}
\def\araa #1 #2{{ARA\&A,}~{#1}, {#2}}
\def\aj #1 #2{{AJ,}~{#1}, {#2}}
\def\alett #1 #2{{\it Astrophys. Lett.,}~{#1}, {#2} }
\def\apj #1 #2{{ApJ,}~{#1}, {#2}}
\def\apjs #1 #2{{ApJS,}~{#1}, {#2}}
\def\aplc #1 #2{{Ap. Lett. Comm.,}~{#1}, {#2}}
\def\apss #1 #2{{Ap\&SS,}~{#1}, {#2}}
\def\astrnach #1 #2{{Astron. Nach.,}~{#1}, {#2}}
\def\azh #1 #2{{AZh,}~{#1}, {#2}}

\def\baas #1 #2{{\it Bull. Am. astr. Soc.,}~{#1}, {#2} }

\def\ca #1 #2{{Comm. Astrophys.,}~{#1}, {#2}}
\def\cpc #1 #2{{Comput. Phys. Comm.,}~{#1}, {#2}}
\def\fcp #1 #2{{\it Fundam. Cosmic Phys.,}~{#1}, {#2} }
\def\jmp #1 #2{{J. Math. Phys.,}~{#1}, {#2}}
\def\jqsrt #1 #2{{J. Quant. Spectrosc. Rad. Transf.,}~{#1}, {#2}}
\def\memsait #1 #2{{\it Memorie Soc. astr. ital.,}~{#1}, {#2} }
\def\mnras #1 #2{{MNRAS,}~{#1}, {#2}}
\def\qjras #1 #2{{\it Q. Jl R. astr. Soc.,}~{#1}, {#2} }
\def\nat #1 #2{{Nature,}~{#1}, {#2}}
\def\pasj #1 #2{{PASJ,}~{#1}, {#2}}
\def\pasp #1 #2{{\it PASP,}~{#1}, {#2}}

\def\physl #1 #2{{\it Phys.Lett.,}~{#1}, #2}
\def\physrep #1 #2{{Phys. Rep.,}~{#1}, #2}
\def\physscri #1 #2{{Phys. Scripta,}~{#1}, #2}

\def\physreva #1 #2{{Phys. Rev. A,}~{#1}, {#2}}
\def\physrevb #1 #2{{Phys. Rev. B,}~{#1}, {#2}}
\def\physrevd #1 #2{{Phys. Rev. D,}~{#1}, {#2}}
\def\physrevl #1 #2{{Phys. Rev. Lett.,}~{#1}, {#2}}
\def\pl #1 #2{{\it Phys. Lett.,}~{#1}, {#2} }

\def\prsl #1 #2{{\it Proc. R. Soc. London Ser. A,}~{#1}, {#2} }

\def\ptp #1 #2{{Prog. Theor. Phys.,}~{#1}, {#2}}
\def\ptps #1 #2{{Prog. Theor. Phys. Suppl.,}~{#1}, {#2}}
\def\rmp #1 #2{{Rev. Mod. Phys.,}~{#1}, {#2}}
\def\rpp #1 #2{{Rep. Progr. Phys.,}~{#1}, {#2}}
\def\sci #1 #2 {{Science,}~{#1}, {#2}}
\def\sovastr #1 #2{{Sov. Astr.,}~{#1}, {#2}}
\def\sovastrl #1 #2{{Sov.Astr. Lett.,}~{#1}, #2}

\def\ssr #1 #2{{\it Space Sci. Rev.,}~{#1}, {#2} }
\def\va #1 #2{{\it Vistas in Astronomy,}~{#1}, {#2} }

\def\book #1 {{\it ``{#1}'',\ }}

\def\ref{\noindent\hangindent=20pt\hangafter=1}

\oneskip
\centerline{REFERENCES}
\oneskip
\parindent=0pt
\parskip=0pt

\ref
Aaronson, M. 1978, \apj 221 L103


\ref
Baud, B., Habing, H.J., Matthews, H.E., \& Winnberg, A. 1981, \aa 95 156

\ref
Baud, B., Sargent, A.J., Werner, M.W., \& Bentley, A.F. 1985, \apj 292 628

\ref
Boulanger, F., \& P\'erault, M. 1988, \apj 330 964

\ref
Burstein, D., Bertola, F., Buson, L.M., Faber, S.M., \& Lauer, T.R. 1988,
\apj 328 440

\ref
Cataloged Galaxies and Quasars Observed in the IRAS Survey, Version 2,
Jet Propulsion Laboratory, 1989

\ref
Cox, P., Kr\"ugel, E., \& Mezger, P.G. 1986, \aa 155 380

\ref
Cox, P., \& Mezger, P.G. 1989, \aar 1 49

\ref
de Vaucouleurs, G., \& de Vaucouleurs, A. 1972, Mem. R. astr. Soc., 77, 1

\ref
de Vaucouleurs, G., de Vaucouleurs, A., Corwin, H.G. Jr., Buta, R.J.,
Paturel, G., \& Fouqu\'e, P. 1991, Third Reference Catalogue of Bright
Galaxies (New York: Springer-Verlag)

\ref
Engels, D., Kreysa, E., Schultz, G.V., \& Sherwood, W.A. 1983, \aa  124 123

\ref
Feast, M.W. 1963, \mnras 125 367

\ref
Feigelson, E.D., \& Nelson, P.I. 1985, \apj 293 192

\ref
Frogel, J.A., \& Whitford, A.E. 1987, \apj 320 199

\ref
Ghosh, S.K., Drapatz, S., \& Peppel, U.C. 1986, \aa  167 341

\ref
Hawarden, T.G., Mountain, C.M., Leggett, S.K., \& Puxley, P.J. 1986,
\mnras 221 41

\ref
Helou, G., Khan, I.R., Malek, L., \& Boehmer, L. 1988, \apjs 68 151

\ref
Herman, J., \& Habing, H.J. 1985, \physrep 124 255

\ref
Impey, C.D., Wynn-Williams, C.G., \& Becklin, E.E. 1986, \apj 309 572

\ref
Johnson, H.L. 1966, \araa 4 193

\ref
Jura, M. 1982, \apj 254 70

\ref
Jura, M., Kim, W., Knapp, R.G., \& Guhathakurta, P. 1987, \apj 312 L11

\ref
King, I.R. 1966, \aj 71 64

\ref
Knapp, G.R., Guhathakurta, P., Kim, D.-W., \& Jura, M. 1989, \apjs 70 329

\ref
Knapp, G.R., Gunn, J.E., \& Wynn-Williams, C.G. 1992, \apj 399 76

\ref
Leisawitz, D., \& Hauser, M.G. 1988, \apj 332 954

\ref
Marsiaj, P. 1992, thesis, Univ. of Padua

\ref
Mazzei, P., Xu, C., \& De Zotti, G. 1992, \aa 256 45

\ref
Munn, J.A. 1992, \apj 399 444

\ref
Paley, E.S., Low, F.J., McGraw, J.T., Cutri, R.M., \& Rix, H.-W. 1991,
\apj 376 335

\ref
Puget, J.L., \& L\'eger, A. 1989, \araa  27 161

\ref
Puget, J.L., L\'eger, A., \& Boulanger, F. 1985, \aa 142 L19

\ref
Puxley, P.J., Hawarden, T.G., \& Mountain, C.M. 1988, \mnras 231 465


\ref
Sandage, A., \& Tammann, G. 1987, A Revised Shapley-Ames Catalog of
Bright Galaxies (Washington: Carnegie Institute)

\ref
Schmitt, J.H.M.M. 1985, \apj 193 178

\ref
Soifer, B.T., Rice, W.L., Mould, J.R., Gillett, F.C., Rowan Robinson, M.,
\& Habing, H.J. 1986, \apj 304 651

\ref
Thronson, H.A. Jr., \& Bally, J. 1987, \apj 319 L63

\ref
Waters, L.B.F.M., Cot\'e, J., \& Aumann, H.H. 1987, \aa 172 225

\ref
Xu, C., \& De Zotti, G. 1989, \aa 225 12

\vfill\eject

\centerline{TABLE 1}

\centerline{MEDIAN FAR-IR TO B FLUX DENSITY RATIOS FOR E AND S0 GALAXIES}

\centerline{$(B_T \leq 12)$}

$$\vbox{\halign{#\hfil&\qquad #\hfil&\quad\hfill# &\qquad\qquad
#\hfil&\quad\hfill#\cr
\noalign{\hrule}&\omit&\omit&\omit&\omit\cr
\noalign{\hrule}&\omit&\omit&\omit&\omit\cr
&\omit&\omit&\omit&\omit\cr
   &\multispan2\hfill E\rlap* \hfill&\multispan2\hfill S0\rlap* \hfill\cr
&\omit&\omit&\omit&\omit\cr
\noalign{\hrule}&\omit&\omit&\omit&\omit\cr
$\log(f_{12}/f_B)\ldots\ldots$ & $\phantom{-}0.01 \pm 0.05\phantom{5}$ &
$(29/47)$ & $-0.06 \pm 0.05\phantom{5}$ & $(33/60)$\cr
$\log(f_{25}/f_B)$\dotfill & $-0.70 \pm 0.32
\phantom{5}$ & $(\phantom{5}8/47)$ & $-0.25 \pm 0.09\phantom{5}$ & $(23/60)$\cr
$\log(f_{60}/f_B)$\dotfill & $-0.22 \pm 0.155$ &
$(15/47)$ & $\phantom{-}0.40 \pm 0.13\phantom{5}$ & $(45/60)$\cr
$\log(f_{100}/f_B)$\dotfill & $\phantom{-}0.25 \pm
0.10\phantom{5}$ & $(23/47)$ & $\phantom{-}0.80 \pm 0.12$ & $(41/60)$\cr
&\omit&\omit&\omit&\omit\cr
\noalign{\hrule}&\omit&\omit&\omit\cr
\multispan5* In parenthesis is the number of detections over the total\hfil
\cr}}$$

\bigskip\bigskip\bigskip

\centerline{TABLE 2}

\centerline{MEDIAN FAR-IR FLUX DENSITY RATIOS FOR E AND S0 GALAXIES}

\centerline{$(B_T \leq 12)$}

$$\vbox{\halign{#\hfil&\qquad #\hfil&\qquad\qquad
#\hfil\cr
\noalign{\hrule}&\omit&\omit\cr
\noalign{\hrule}&\omit&\omit\cr
&\omit&\omit\cr
   &\hfill E \hfill&\hfill S0 \hfill\cr
&\omit&\omit\cr
\noalign{\hrule}&\omit&\omit\cr
$\log(f_{12}/f_{100})\ldots\ldots$ & $-0.35 \pm 0.20\phantom{5}$ &
$-0.95 \pm 0.17$ \cr
$\log(f_{25}/f_{100})$\dotfill & $-0.8\phantom{5} \pm
0.4\phantom{5}\phantom{5}$ & $-1.10 \pm 0.16 $ \cr
$\log(f_{60}/f_{100})$\dotfill & $-0.45 \pm 0.15\phantom{5}$ &
$-0.40 \pm 0.05\phantom{5}$ \cr
&\omit&\omit\cr
\noalign{\hrule}&\omit\cr}}$$

\vfill\eject

\vfill\eject

\centerline{\bf Figure captions}

\bigskip\noindent
{\bf Fig. 1.} Distributions of logarithms of far-IR to $B_T(0)$ flux
density ratios of the 47 E galaxies brighter than $B_T = 12$ in the RSA
catalog, for the four IRAS bands,
reconstructed exploiting the Kaplan-Meyer estimator, taking into
account upper limits (solid lines). Dashed are the distribution of
galaxies {\it detected} in the relevant IRAS band; there are 29 galaxies
detected at $12\mic$, 8 detected at $25\mic$, 15 detected at $60\mic$,
23 detected at $100\mic$.

\bigskip\noindent
{\bf Fig. 2.} Same as in Fig. 1, but for S0 galaxies (60 objects);
33 are detected at $12\mic$, 23 at $25\mic$, 45 at $60\mic$,
41 at $100\mic$.

\bigskip\noindent
{\bf Fig. 3.} Same as in Fig. 1, but for E galaxies brighter than
$B_T = 16$ (513 objects) in the sample by Knapp et al. (1989). Note the
extended tail
towards high far-IR fluxes, likely due to misclassified galaxies.

\bigskip\noindent
{\bf Fig. 4.} Distribution of $60/100\mic$ and $25/100\mic$ flux density
ratios for elliptical (left) and S0 galaxies. On the right hand panels,
the point corresponding to the galaxy NGC~5192 (taken to be an S0, but whose
morphological classification is uncertain) does not appear because
its flux density ratios ($\log f_{25}/f_{100} > 1.57$ and
$\log f_{60}/f_{100} > 1.45$) fall out of the frames.

\bigskip\noindent
{\bf Fig. 5.} Distribution of $12/2.2\mic$ against $12/100\mic$ flux density
ratios of elliptical [panel a)] and S0 galaxies [panel b)].

\bigskip\noindent
{\bf Fig. 6.} Average far-IR spectral energy distributions, normalized
to the $60\mic$ flux, of E (circles) and S0 (asterisks) galaxies.
Also shown, for comparison, are the
far-IR spectra of nearby disk galaxies (squares) and of non-Seyfert
Markarians (triangles), as estimated by Xu \& De Zotti (1989).

\bye